\documentclass[sigconf]{acmart}

\usepackage[T1]{fontenc}
\usepackage{booktabs} 
\usepackage{xcolor}
\usepackage{amsmath}
\usepackage{balance}

\definecolor{temp}{HTML}{FF0000}
\definecolor{todo}{HTML}{FFA419}
\definecolor{done}{HTML}{0BF9AB}
\definecolor{info}{HTML}{1E90FF} 
\definecolor{moresamples}{HTML}{00FF00}

\begin{document}

\copyrightyear{2018}
\acmYear{2018}
\setcopyright{acmlicensed}
\acmConference[UMAP '18]{26th Conference on User Modeling, Adaptation and Personalization}{July 8--11, 2018}{Singapore, Singapore}
\acmBooktitle{UMAP '18: 26th Conference on User Modeling, Adaptation and Personalization, July 8--11, 2018, Singapore, Singapore}
\acmPrice{15.00}
\acmDOI{10.1145/3209219.3209236}
\acmISBN{978-1-4503-5589-6/18/07}

\fancyhead{}

\title{Multi-Modal Adversarial Autoencoders for Recommendations \\ of Citations and Subject Labels}
\renewcommand{\shorttitle}{Multi-Modal Adversarial Autoencoders for Recommendations}

\author{Lukas Galke}
\orcid{0000-0001-6124-1092}

\affiliation{
  \institution{Kiel University}
  \country{Germany}
}
\email{lga@informatik.uni-kiel.de}

\author{Florian Mai}

\affiliation{
  \institution{Kiel University}
  \country{Germany}
}
\email{stu96542@informatik.uni-kiel.de}

\author{Iacopo Vagliano}
\affiliation{
  \institution{ZBW -- Leibniz Information Centre for Economics}
  \city{Kiel}
  \country{Germany}
}
\email{I.Vagliano@zbw.eu}

\author{Ansgar Scherp}
\affiliation{
  \institution{Kiel University}
  \country{Germany}
}
\email{asc@informatik.uni-kiel.de}

\renewcommand{\shortauthors}{L. Galke, F. Mai, I. Vagliano, A. Scherp}

\begin{abstract}

We present multi-modal adversarial autoencoders for recommendation and evaluate
them on two different tasks: citation recommendation and subject label
recommendation.  We analyze the effects of adversarial regularization, sparsity,
and different input modalities.  By conducting 408 experiments, we show that
adversarial regularization consistently improves the performance of autoencoders
for recommendation.  We demonstrate, however, that the two tasks differ in the
semantics of item co-occurrence in the sense that item co-occurrence resembles
relatedness in case of citations, yet implies diversity in case of subject
labels.  Our results reveal that supplying the partial item set as input is only
helpful, when item co-occurrence resembles relatedness.  When facing a new
recommendation task it is therefore crucial to consider the semantics of item
co-occurrence for the choice of an appropriate model.

\end{abstract}




\maketitle

\newcommand{\Xtrain}{X_\text{train}}
\newcommand{\train}{\text{train}}
\newcommand{\test}{\text{test}}
\newcommand{\pred}{\text{pred}}

\newcommand{\Xtest}{X_\text{test}}
\newcommand{\Sim}[2]{\operatorname{sim}\left(#1, #2\right)}
\newcommand{\SimOp}{\Sim{\cdot}{\cdot}}
\newcommand{\ie}{i.\,e.}
\newcommand{\mlp}{\operatorname{MLP-2}}
\newcommand{\Nexperiments}{408}

\section{Introduction}

Recent advances in autoencoders on images have shown that adversarial regularization can improve the performance of autoencoders~\cite{DBLP:journals/corr/MakhzaniSJG15}.
The so-called adversarial autoencoders~\cite{DBLP:journals/corr/MakhzaniSJG15} are not only trained to reconstruct the input, but also to match the code with a selected prior distribution.
We hypothesize that the thereby imposed smoothness on the code aids autoencoders in reconstructing highly sparse item vectors for recommendation.
The rationale is that smoothness is one of the criteria for good representations that disentangle the explanatory factors of variation~\cite{DBLP:journals/corr/abs-1206-5538}.
In this paper, we analyze whether adversarial autoencoders can be applied to highly sparse recommendation tasks.
We evaluate the effect of adversarial regularization with respect to the degree of sparsity and different input modalities on two exemplary tasks: citation and subject label recommendation.

\begin{description}
\item[Citation Recommendation]
More and more publishers decide to contribute to the Initiative for Open
Citations\footnote{\url{https://i4oc.org}}, which aims to make citation metadata
publicly available. This motivates us to consider the following scenario as a
recommendation task. When writing a new paper, it is essential that the
authors reference other publications which are key in the respective field of
study or relevant to the paper being written. Failing to do so can be rated
negatively by reviewers in a peer-reviewing process. However, due to increasing
volume of scientific literature, even some critical paper are sometimes 
overlooked. Hence, in this paper we study the problem of recommending
publications to consider as citation candidates, given that the authors have
already selected some other references and assuming that the paper is close to
completion, \ie{}, information such as the title (or a tentative title) of the paper
is available.

\item[Subject Indexing]
Apart from citation data, also subject labels, or tags,
are publicly available for numerous domains, such as
MeSH\footnote{\url{https://www.nlm.nih.gov/mesh/}} for medicine or
EconBiz\footnote{\url{https://www.econbiz.de/}} for economics.
Subject indexing is a common task in scientific libraries to
make documents accessible for search. New documents are annotated with a
set of subjects by professional subject indexers. Fully-automated multi-label
classification approaches to subject indexing are
promising~\cite{DBLP:conf/nips/NamMKF17}, even when merely the metadata of the
publications is used~\cite{DBLP:conf/kcap/GalkeMSBS17}. Professional subject
indexers, however, typically use the result of these approaches only as recommendations,
so that the human-level quality can still be guaranteed. This circumstance
motivates us to build a subject label recommender system that explicitly takes the partial
list of already assigned subjects into account.
\end{description}

To unify these two scenarios, we take either the citations or the assigned subjects as implicit feedback for
a considered recommendation task.
In the former case, citations are known to
resemble credit assignment~\cite{wouters1999citation},
whereas in the latter case the subject labels are selected by respective
professionals such that their relevance to the paper is guaranteed by human
supervision.

Traditionally, the recommendation problem is modeled as the prediction of missing
ratings in a \(\boldsymbol U \times \boldsymbol I\) matrix with set of users \(\boldsymbol
U\) and set of items \(\boldsymbol I\) (matrix
completion). In our case, following
\citeauthor{McNee:2002:RCR:587078.587096}~\cite{McNee:2002:RCR:587078.587096}, we
view research papers themselves as users over their authors or the responsible subject
indexers. The rationale is that one author may be involved in multiple papers
of different domains but that all authors for a given paper should receive the same recommendations.
Analogously, a given paper should receive the same recommendations for candidate
subjects, independently of the current subject indexer in charge of annotating it.

We have transferred the approach of
\citeauthor{DBLP:journals/corr/MakhzaniSJG15}~\cite{DBLP:journals/corr/MakhzaniSJG15},
which was applied to images, and extended it to our problem of a general
recommendation task.
By developing a novel interpretation of the adversarial autoencoder, we show
how it can be applied to recommendation tasks and how multiple input modalities can be
incorporated.
We make use of this capability in our experiments by considering besides the
ratings also additional metadata, namely the documents' title, as content-based
features.
We performed \Nexperiments{} experiments for our two recommendation
tasks to study how
adversarial autoencoders perform while exploiting titles along
with the partial list of citations or the already assigned subjects,
respectively. For a close investigation of the adversarial autoencoders' performance, we not only
consider the adversarial autoencoder as a whole but also individually assess its components.

We further evaluate to which degree these models are robust to sparsity in the dataset.
When conducting citation or research paper recommendation, it is not
desirable that only already frequently cited papers get recommended and less
frequently cited papers are ignored. Common pruning strategies
comprise removing rarely cited documents and documents that cite too few other
works~\cite{beel2016paper}. This pruning step affects the number of
considered items, and thus, the degree of sparsity. To gain a better
understanding of how the pruning threshold affects the models'
performance, we conduct experiments, in which the pruning threshold is a
controlled variable.

Our results show that the partial list of items is more important for the
citation recommendation task than it is for the subject labeling task.
This is interesting because an inspection of the semantics of item co-occurrence
may help researchers or practitioners to tackle new recommendation tasks,
specifically to decide whether to supply the partial list of items as input.
For citation recommendation, item co-occurrence implied relatedness, \ie{}, it
is of
high relevance which other works have been cited so
far. For subject labels, in contrast, co-occurrence implies diversity:
similar subjects are rarely used together for annotation of a single document.
Thus, the title is more relevant than the already assigned subjects.
All of the evaluated methods appeared similarly sensitive to data
sparsity despite the differences in the number of parameters.

Due to the use of the titles, the adversarial autoencoders yields competitive
performance to the baselines. On the subject label recommendation task, they
outperform the baselines. A closer look at the individual components of the
adversarial autoencoder revealed that the sole MLP decoder achieved better
performance than the whole model on the subject labelling task, while its
performance fell behind the whole model on the citation recommendation task.

In summary, our contributions are the following:
\begin{itemize}
  \item We present an adaption of adversarial autoencoders as a novel approach
  for multi-modal recommendation tasks on scientific documents.
  \item We analyze this new method along with all of its components on citation and
  subject label recommendation tasks while varying the input modalities.
  We gain valuable insights on the interactions between input modalities and the
  task: when item co-occurrence resembles relatedness, multi-modal variants are preferable,
  otherwise solely content-based variants may be more suitable.
  \item We evaluate the autoencoder models in realistic scenarios, as we split
  the datasets on the time axis and consider different thresholds for pruning by
  minimum item occurrence. This is especially important for the citations task
  because only already existing papers can be cited and it is desirable that also less cited papers are recommended.
\end{itemize}

The remainder of this paper is organized as follows. In Section~\ref{sec:rw}, we
review previous work on citation and tag recommendation as well as
recommendation approaches from the deep learning field. After formally stating
the problem in Section~\ref{sec:problem-statement}, we introduce the employed
models in Section~\ref{sec:models}, describe the citation and subject
recommendation experiments in Section~\ref{sec:experiments}. We discuss the
results in Section~\ref{sec:discussion}, before we conclude.

\section{Related work}\label{sec:rw}
\paragraph{Research paper and subject label recommendation}
An extensive survey~\cite{beel2016paper} shows that research paper
recommendation is a well-known topic. 
In this context, BibTip~\cite{geyer2002recommendations} and bX~\cite{DBLP:conf/jcdl/BollenS06}
are well-known recommender systems, which operate on the basis of citations
harvested by CiteSeer~\cite{DBLP:conf/dl/GilesBL98}. Docear is a more recent
research paper recommender system, which takes user profiles into
account~\cite{DBLP:journals/dlib/BeelLGN14}.
For citation recommendation specifically,
\citeauthor{DBLP:conf/cikm/HuangKCMGR12} distinguish between recommendations
based on a partial list of references and recommendations based on the content
of a manuscript~\cite{DBLP:conf/cikm/HuangKCMGR12}. While the former is suited
for finding matching citations for a given statement during writing, the latter
strives to identify missing citations on the broader, document level.
Citation recommendation recently focuses on context-sensitive applications,
in which concrete sentences are mapped to, preferably relevant,
citations~\cite{beel2016paper,DBLP:conf/cikm/HuangKCMGR12,DBLP:conf/sigir/EbesuF17}.
Instead, we revisit the reference list completion problem
and we do not take into account the context of the citation, as the full text of
a papers is rarely available in large-scale metadata sources. 
In 1973, \citeauthor{DBLP:journals/jasis/Small73} started the field of
co-citation analysis~\cite{DBLP:journals/jasis/Small73}.
Co-citation analysis assumes that two papers are more related to each other, the
more they are co-cited.
Following that idea, \citeauthor{caragea2013can} relied on singular value
decomposition as a more efficient and extendable approach~\cite{caragea2013can}.
We recognize the need for new methods that are not only based on
item co-occurrence but also take supplementary metadata into account for these
partial list completion problems.

Subject label recommendation is similar to tag recommendation, as in both cases
the goal is to suggest a descriptive label for some content.
\citeauthor{Sen:2009:TCU:1526709.1526800} propose algorithms that predict users'
preferences for items based on their inferred preferences for
tags~\cite{Sen:2009:TCU:1526709.1526800}.
\citeauthor{Montanes:2009:CTR:3056147.3056161} exploit probabilistic regression
for collaborative tag recommendation~\cite{Montanes:2009:CTR:3056147.3056161},
while \citeauthor{Krestel:2009:LDA:1639714.1639726} relied on
Latent Dirichlet allocation~\cite{Krestel:2009:LDA:1639714.1639726}. Similarly, \citeauthor{Sigurbjornsson:2008:FTR:1367497.1367542} propose a tag recommender for Flickr
to support the user in the photo annotation
task~\cite{Sigurbjornsson:2008:FTR:1367497.1367542},
whereas \citeauthor{Posch:2013:MCU:2487788.2488008} predict hashtag categories
on Twitter~\cite{Posch:2013:MCU:2487788.2488008}.
\citeauthor{Dellschaft:2012:MIT:2309996.2310009} measure the influence of tag
recommender systems on the indexing quality in collaborative tagging systems~\cite{Dellschaft:2012:MIT:2309996.2310009}.
These works, however, focus on tags for social media, while we
consider subject labels from a standardized thesaurus for scientific documents.

\paragraph{Recommendation and Link Prediction based on Deep Learning}
Multiple recommender systems based on deep learning have been proposed. \citeauthor{DBLP:conf/kdd/WangWY15} used
deep learning for collaborative filtering~\cite{DBLP:conf/kdd/WangWY15}.
Another recent collaborative-filtering approach explicitly takes side
information into account for
autoencoders~\cite{DBLP:journals/eswa/BarbieriABZ17}. We include a
similar model in our comparison, as it is one component of the adversarial autoencoder.
Additional techniques employ recurrent neural networks to provide session-based
recommendations~\cite{DBLP:conf/recsys/QuadranaKHC17} or
combine knowledge graphs with deep
learning~\cite{DBLP:conf/recsys/Palumbo0T17,DBLP:conf/recsys/RosatiRNLP16}.
To the best of our knowledge, only two approaches makes use of deep learning techniques for citation
recommendation. However, both of them focus on context-sensitive
scenarios~\cite{Huang:2015:NPM:2886521.2886655,DBLP:conf/sigir/EbesuF17}.

Citation networks are also considered in many studies on link prediction.
By making use of the network structure, dedicated architectures learn
representations of
its nodes. One of the most prominent approaches is
DeepWalk~\cite{DBLP:conf/kdd/PerozziAS14}, together with its extension
Node2vec~\cite{node2vec-kdd2016}. These methods perform a random walk over the
graph and feed the generated sequence into skip-gram negative sampling methods~\cite{DBLP:conf/nips/MikolovSCCD13}.
\citeauthor{DBLP:journals/corr/KipfW16} recently proposed Graph
Auto-Encoders~\cite{DBLP:journals/corr/KipfW16a} and Graph Convolutional
Networks~\cite{DBLP:journals/corr/KipfW16}. However, all of these graph-based
approaches assume that all nodes (research papers) are known
during training.
Hence, they are unable to deal with unknown nodes (new, unseen citing
documents) at test time. Instead, we focus on a more realistic application scenario,
where we need to predict citations for a paper which is being written and thus
yet unknown. To simulate such practical settings, we ensure that all
documents of the test set are unknown to the system during training.
Such a scenario is challenging as it corresponds to a cold-start situation.

\section{Problem Statement}\label{sec:problem-statement}
\begin{table}
  \caption{Notation}\label{tab:notation}
\begin{tabular}{ll}
  \toprule
  Symbol & Description \\
  \midrule
  \(\mathbb D\) & Set of \(m\) documents \\
  \(\mathbb I\) & Set of \(n\) items \\
  \(\boldsymbol{X} \in {\lbrace 0,1 \rbrace}^{m \times n}\) & Sparse ratings matrix \\
  \(\boldsymbol{S} \in {\mathbb{R}}^{m \times d}\) & Supplementary document
  information \\
  \(\boldsymbol{x}, \boldsymbol{s} \) & Row vectors of \( \boldsymbol{X} \) or \( \boldsymbol{S} \), respectively \\
  \(\lbrack \boldsymbol{x}; \boldsymbol{s} \rbrack \) & Concatenation of vectors \(\boldsymbol{x} \) and \(\boldsymbol{s} \) \\
  \(\bowtie \) & Natural join (on document identifiers) \\
  \( \boldsymbol{I} \) &  Identity matrix \\
  \bottomrule
\end{tabular}
\end{table}

In the following, we provide a formal problem statement for the considered
recommender task. The documents can be considered users in a traditional
recommendation scenario, while the items are either cited documents or subject
labels, respectively.

Given a set of \( m \) documents \(\mathbb D\) and
a set of \( n \) items \(\mathbb I\), the typical recommendation task is to model
the spanned space, \(\mathbb D \times \mathbb I\).
We model the ratings as a sparse matrix \( \boldsymbol{X} \in { \lbrace 0,1 \rbrace } ^ {m \times
n}\), in which \( X_{jk} \) indicates implicit feedback from document \( j \) to item \( k \).
To simulate a real-world scenario, we split
the documents \( \mathbb{D} \) into \(m_\train \) documents for training \(\mathbb{D}_\train \) and \( m_\test \) documents for evaluation \(\mathbb{D}_\test \),
such that \(\mathbb{D}_\train \cap \mathbb{D}_\test =
\emptyset \). 
More precisely, we conduct this split into training and test documents based on
the publication year. All documents that were published before a certain year
are used as training, and the remaining documents as test data.
This leads to an experimental setup that is close to a real-world application.
More details will be provided in Section~\ref{sub:citations}.
All models are supplied with the complete ratings of the users
\(\boldsymbol{X}_\train = \mathbb{D}_\train \bowtie \boldsymbol{X}\) along
with the supplementary information \( \boldsymbol{S}_\train =
\mathbb{D}_\train \bowtie \boldsymbol{S}\) for training.
In the present work, we use the title of the documents as supplementary
information. Still, in theory, more sources of supplementary information may be considered.
The test set \(\boldsymbol{X}_\test, \boldsymbol{S}_\test \) is obtained analogously.

For evaluation, we remove randomly selected items in \( \boldsymbol{X}_\test \) by setting one
non-zero entry in each row to zero. We denote the hereby created test set by \( \boldsymbol{\tilde X}_\test \).
The model ought to predict values \(\boldsymbol{X}_\pred \in {\lbrack 0,1
\rbrack}^{m_\test \times n} \), given the test set \( \boldsymbol{\tilde X}_\test \)
along with the title information \( \boldsymbol{S}_\test \).
Finally, we compare the predicted ratings \(\boldsymbol{X}_\pred \) with the
true ratings \( \boldsymbol{X}_\test \) via ranking metrics.
The goal is that those items, that were omitted in 
\( \boldsymbol{\tilde X}_\test \), are highly ranked in \( \boldsymbol{X}_\pred \).

\begin{figure}
  \includegraphics[height=5.5cm,keepaspectratio]{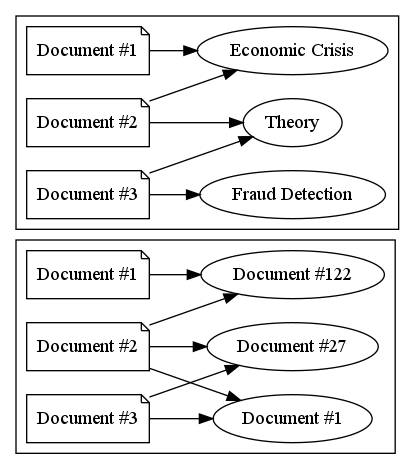}
  \caption{Exemplary bipartite graphs of documents annotated with subject labels
    (top) and citation relationships between documents (bottom). It becomes
  apparent how the two recommendation tasks share a similar structure.}\label{fig:example-viz}
\end{figure}

In both scenarios, \ie{}, citation recommendation and subject label recommendation, we
regard documents and items as a bipartite graph (see
Figure~\ref{fig:example-viz}). Considering citations, this point of view may be
counter-intuitive since a scientific document is typically both a citing paper and a cited
paper. Still, the out-degree of typical citation network datasets is so high
that we cannot expect to have metadata for all cited papers available. For
instance, the PubMed citation dataset we use for our experiments offers
metadata of 224,092 documents that cite 2,896,764 distinct other documents.
Therefore it is reasonable to rely only on the metadata of the citing documents
itself as basis for recommendations.

\section{Models}\label{sec:models}
In the following, we describe the employed models. We start with two baselines
based on item co-occurrence. Subsequently, we briefly introduce the multi-layer
perceptron as a building block for the two autoencoder variants. We show how
title information can be incorporated in undercomplete and adversarial
autoencoders. We provide information on hyperparameters
in the final paragraph of this section.

\paragraph{Item Co-Occurrence}\label{methods:baseline:cocit} As a non-parametric
yet strong baseline we consider the co-citation
score~\cite{DBLP:journals/jasis/Small73} that is purely based on item co-occurrence.
The rationale is that two papers, which have been cited more often
together in the past, are more likely to be cited together in the future than
papers that were less often cited together.
Given training data \(\boldsymbol{X}_\train \), we compute the full item
co-occurrence matrix \(\boldsymbol{C} = {\boldsymbol{X}_\train}^T \cdot
\boldsymbol{X}_\train \in \mathbb{R}^{n \times n}\). At prediction time,
we obtain the scores by aggregating the co-occurrence values via matrix
multiplication \(\boldsymbol{X}_\test \cdot \boldsymbol{C}\). On the diagonal of
\( \boldsymbol{C} \), the (squared) occurrence count of each item is retained to
model the prior probability.

\paragraph{Singular Value Decomposition} Singular value decomposition (SVD) is
an approach that factorizes the co-occurrence matrix of items \( \boldsymbol{X}^T \cdot \boldsymbol{X} \).
\citeauthor{caragea2013can} showed that SVD can be successfully used for citation
recommendation~\cite{caragea2013can}. We therefore include SVD in our comparison and
extend it by the capability of incorporating title information, which has
already been proposed as future work by
\citeauthor{caragea2013can}~\cite{caragea2013can}. We concatenate the textual
features as TF-IDF weighted bag-of-words with the items and perform singular value
decomposition on the resulting matrix.
To obtain predictions, we only use those
indices of the reconstructed matrix that are associated with items.

\paragraph{Multi-Layer Perceptron}\label{sec:dist-rep-mlp}

A multi-layer perceptron (MLP) is a fully-connected feed-forward neural network
with one or multiple hidden layers. The output is computed by consecutive
applications of \(\boldsymbol{h}^{(i)} = f(\boldsymbol{h}^{(i-1)} \cdot
\boldsymbol{W}^{(i)} + \boldsymbol{b}^{(i)})\) with \(f\) being a nonlinear
activation function. 
In the description of the following models, we
abbreviate a two hidden-layer perceptron module by MLP-2. This MLP-2 module is not
only used as a building block for subsequent architectures, but also  as a full
model that only operates on the documents' titles.
In this case, we optimize binary cross-entropy \( \operatorname{BCE}( x,
\operatorname{MLP-2}(s) ) \), where the titles \( s \) are used as input and
citations or subject labels \( x \) as target outputs.
We chose to operate on an TF-IDF weighted embedded bag-of-words
representation~\cite{DBLP:conf/gi/GalkeSS17}
for a fair comparison with the autoencoder variants, which are described below.

\paragraph{Undercomplete Autoencoders} The general concept of an autoencoder (AE)
involves two components: the encoder \texttt{enc} and the decoder
\texttt{dec}. The encoder transforms the input into a hidden
representation (the code) \( z = \operatorname{enc}(x) \). Then the decoder
reconstructs the input from the code \(r = \operatorname{dec}(z)\).
The two components are jointly trained to minimize 
the loss function \(BCE(x,r)\). To avoid learning to merely copy the input \(x\) to
the output \(r\), autoencoders need to be regularized. The most common
way to regularize autoencoders is by imposing a lower dimensionality on the code
(undercomplete). In short, autoencoders are 
trained to capture the most important explanatory factors of variation for
reconstruction~\cite{DBLP:journals/corr/abs-1206-5538}.

For both the encoder and the decoder we chose an MLP-2 module, such that the model function becomes \(
r = \operatorname{MLP-2_{dec}}(\operatorname{MLP-2_{enc}}(x)) \). When the
documents' title is available, we supply it as additional input to the decoder
\( r = \operatorname{MLP-2_{dec}}(\lbrack
\operatorname{MLP-2_{enc}}(x); s\rbrack) \). We embed the textual features into
a lower dimensional space by using pre-trained word embeddings~\cite{DBLP:conf/nips/MikolovSCCD13}.
The rationale here is that the rather low code dimension is not overwhelmed by
the high amount of vocabulary terms. For a fair comparison of the models,
also the MLP described above is supplied the same text representation as input.
More precisely,
we employ a TF-IDF weighted bag of embedded words representation which has
proven
to be useful for information retrieval~\cite{DBLP:conf/gi/GalkeSS17}.
The usage of title information in an undercomplete autoencoder is comparable to the
approach by
\citeauthor{DBLP:journals/eswa/BarbieriABZ17}~\cite{DBLP:journals/eswa/BarbieriABZ17}.
A minor difference is that we supply the side information (titles) only to the
decoder, yet use two hidden layers for both the encoder and the decoder to
enable a fair comparison to the adversarial variant, which is described below.

\begin{figure*}
  \includegraphics[width=\textwidth]{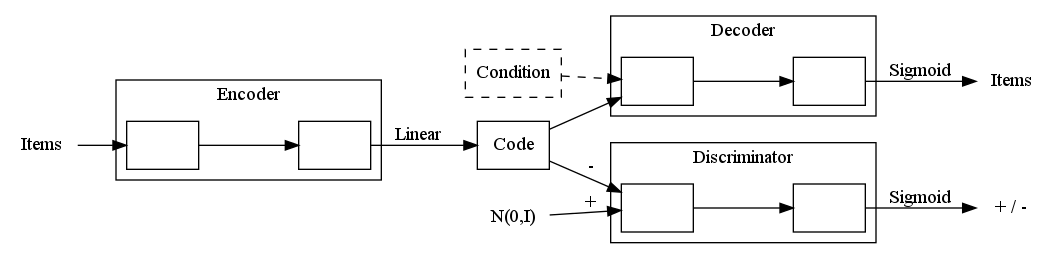}
  \caption{Adversarial autoencoder for item-based recommendations. Each edge
    resembles a parametrized mapping \(f(\boldsymbol{W} \boldsymbol{x} + \boldsymbol{b})\) with activation function \(f\) and
  parameters \(W, b\). When not labeled differently, the activation
  function is rectified linear followed by dropout.}\label{fig:aae}
\end{figure*}

\paragraph{Adversarial Autoencoders} We extend the work of
\citeauthor{DBLP:journals/corr/MakhzaniSJG15} on adversarial
autoencoders (AAE)~\cite{DBLP:journals/corr/MakhzaniSJG15}, who combine generative
adversarial networks~\cite{DBLP:conf/nips/GoodfellowPMXWOCB14} with
autoencoders. The autoencoder component reconstructs the sparse item
vectors, while the discriminator distinguishes between the generated codes and
samples from a selected prior distribution (see Figure~\ref{fig:aae}). Hence,
the distribution of the latent code is shaped to match the prior distribution.
We hypothesize that the latent representations learned by distinguishing the code
from a smooth prior lead to a model that is more robust to sparse input
vectors than undercomplete autoencoders. The rationale is that smoothness is a
main criterion for good representations that disentangle the explanatory factors
of variation~\cite{DBLP:journals/corr/abs-1206-5538}.

Formally, we first compute \( h = \mlp_\text{enc}(x) \)
and \( r = \mlp_\text{dec}(h) \) and then update the parameters of the encoder
and the decoder with respect to binary cross-entropy \( \operatorname{BCE}(x, r) \).
Hence, in the regularization phase, we draw samples \( z \sim \mathcal{N}(0,
\boldsymbol{I}) \) from independent Gaussian distributions matching the size of
\( h \). The parameters of the discriminator \( \mlp_\text{disc} \) are then
updated, to minimize \( \log \mlp_\text{disc}(z) + \log (1 -
\mlp_\text{disc}(h)) \)~\cite{DBLP:conf/nips/GoodfellowPMXWOCB14}.
Finally, the parameters of the encoder are updated to
maximize \( \log \mlp_\text{disc}(h) \), such that the encoder is trained to
fool the discriminator. As a result, the encoder is jointly optimized for
matching the prior distribution and for reconstruction of the
input~\cite{DBLP:journals/corr/MakhzaniSJG15}.

To incorporate the documents' title, we once again concatenate on the code
level. This scenario corresponds to the supervised case from the original work of
\citeauthor{DBLP:journals/corr/MakhzaniSJG15} on images, in which the purpose was to
separate the style from the class. All information that cannot be reconstructed
from the class is drawn from the style (the
code)~\cite{DBLP:journals/corr/MakhzaniSJG15}. We adapt this interpretation
by supplying title information as additional input to the decoder.
Hence, the model is optimized to exploit the title information when it is helpful
for reconstruction but also take the partial item set into account.
At prediction time, we perform one reconstruction step by applying one encoding
and one decoding step.

\paragraph{Hyperparameters} The hyperparameters are selected by conducting
pre-experiments on the citation recommendation dataset by considering only items that appear 50
or more times in the whole corpus. We chose this scenario because this
aggressive pruning results in numbers of distinct items and documents that are similar to the
ones of the subject label recommendation dataset. Considering the MLP-modules, we conducted a grid
search with hidden layer sizes between 50 and 1,000, initial learning
rates between \(0.01\) and \(0.00005 \), activation functions Tanh,
ReLU~\cite{DBLP:conf/icml/NairH10}, SELU~\cite{DBLP:conf/nips/KlambauerUMH17}
along with dropout~\cite{DBLP:journals/jmlr/SrivastavaHKSS14} (or alpha-dropout in case of SELUs) probabilities between
\(.1\) and \(.5\) and as optimization algorithms stochastic gradient descent and
Adam~\cite{DBLP:journals/corr/KingmaB14}. For the autoencoder-based models, we
considered code sizes between 10 and 500, but only if the size was
smaller than the hidden layer sizes of the MLP modules. In case of adversarial
autoencoders, we experimented with Gaussian, Bernoulli, and Multinomial prior
distributions, and with linear, sigmoid, and softmax activation on the code layer,
respectively.

While we do not exclude that a certain set of hyperparameters may
perform better in a specific scenario, we select the following, most
robust, hyperparameters:
hidden layer sizes of 100 with ReLU~\cite{DBLP:conf/icml/NairH10}
nonlinearities and drop probabilities of \( .2 \) after each hidden
layer. The optimization is carried out by Adam~\cite{DBLP:journals/corr/KingmaB14}
with initial learning rate \( 0.001 \). The two autoencoder variants use a
code size of 50. We further select a
Gaussian prior distribution for the adversarial autoencoder.
For SVD, we consecutively increased the number of singular values up to 1,000.
Using higher amounts of singular values decreased the performance.
We keep this set of hyperparameters fixed across all models and across
all subsequent experiments to ensure a reliable comparison of the models' quality.

\section{Experiments}\label{sec:experiments}

To evaluate adversarial autoencoders for recommendation tasks on scientific
documents, we conduct a citation recommendation experiment as presented in
Section~\ref{sub:citations} and a subject label recommendation experiment as
presented in Section~\ref{sub:subjects}. Adversarial autoencoders are not only
evaluated against the two baselines (item co-occurrence and SVD), but also
against its own components: undercomplete autoencoders and multi-layer
perceptrons.

\subsection{Citation Recommendation}\label{sub:citations}

In this section, we describe our experimental setup which is designed to
resemble a real-world application of missing citation recommendation.

\paragraph{Dataset}
The CITREC\footnote{\url{https://www.isg.uni-konstanz.de/projects/citrec/}} PubMed
citation dataset~\cite{gipp2015citrec} consists of 7,546,982 citations.
The dataset comprises 224,092 distinct citing documents published between 1928
and 2011 and 2,896,764 distinct cited documents.
The documents are cited between 1 and 3,247 times with a median of 1
and a mean of 2.61 (SD\@: 6.71).
The citing documents hold on average 33.68 (SD\@: 27.49) citations
to other documents (minimum: 1, maximum 2,242) with a median of 29.

\begin{figure}[!ht]
  \includegraphics[width=0.8\columnwidth]{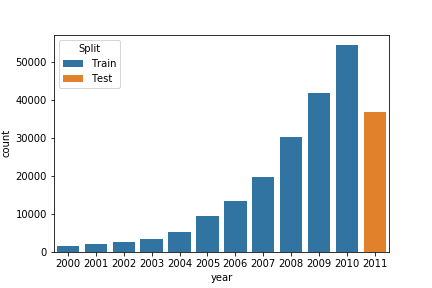}
  \caption{Count of documents by publication year starting with 2000 along with
  the split in training and test set for the PubMed citation dataset.}\label{fig:pmcc_by_year}
\end{figure}

\paragraph{Split on Time Axis} To simulate a real-world citation prediction
setting, we split the data on the time axis of the citing documents. This
resembles the natural constraint that publications cannot cite other
publications that do not exist yet. Given a specific publication year \(T\), we
ensure that the training set \( \boldsymbol{D}_\train \) consists of all documents that were
published earlier than year \(T\) and use the remaining documents as test data
\(\boldsymbol{D}_\test \).
Figure~\ref{fig:pmcc_by_year} shows the distribution of documents over the years
along with the split into training and test set. We select the year 2011 for
evaluation to obtain a 90:10 ratio between training and test documents.

\begin{table}
  \centering
  \caption{Dataset characteristics with respect to pruning thresholds on minimum item
  occurrence for the PubMed citation recommendation task.}\label{tab:pubmed:pruning}
  \begin{tabular}{lrrrr}
\toprule
pruning &  cited documents &  citations &  documents &   density \\
\midrule
15 &    35,664 &    1,173,568 &  136,911 &  0.000240 \\
20 &    20,270 &     878,359 &   121,374 &  0.000357 \\
25 &    12,881 &     692,037 &   105,170 &  0.000511 \\
30 &     8,906 &     568,563 &    96,980 &  0.000658 \\
35 &     6,469 &     478,693 &    87,498 &  0.000846 \\
40 &     4,939 &     413,746 &    79,830 &  0.001049 \\
45 &     3,904 &     363,870 &    73,200 &  0.001273 \\
50 &     3,185 &     324,693 &    67,703 &  0.001506 \\
55 &     2,643 &     292,791 &    62,647 &  0.001768 \\
\bottomrule
\end{tabular}

\end{table}

\paragraph{Preprocessing and Dataset Pruning as Controlled Variable}
For preprocessing the datasets, we conduct the following three steps:
\begin{enumerate}
  \item Build a vocabulary on the training set with items that received implicit
  feedback more than \( \alpha \) times.
  \item Filter both the training and test set and retain only items from the
  vocabulary.
  \item Remove documents that are assigned to fewer than two of the vocabulary
  items.
\end{enumerate}
The pruning threshold \( \alpha \) is crucial since it affects both the number of
considered items as well as the number of documents.
Thus, we identify \( \alpha \) as a controllable parameter and evaluate the models'
performance with respect to different values for \( \alpha \).
Table~\ref{tab:pubmed:pruning} shows the dataset characteristics
with respect to the pruning threshold.

\paragraph{Evaluation Metric} For evaluation, certain items were omitted on
purpose in the test set.
For each document, the models ought to predict the omitted item as good as
possible. Thus, we choose mean
reciprocal rank as our evaluation metric.
We are given a set of predictions
\(\boldsymbol{X}_\text{pred} \) for the test set \( \boldsymbol{\tilde X}_\test \). Hence for each row, we compute the reciprocal rank of
the missing element from \( \boldsymbol{x}_\test - \boldsymbol{\tilde x}_\test \). The
reciprocal rank corresponds to one divided by the position of the omitted item
in the sorted list of predictions \(\boldsymbol{x}_\text{pred} \). We then
average over all documents of the test set to obtain the mean reciprocal rank.
To alleviate random effects of model initialization, training
data shuffling, and selecting the elements to omit, we conduct three runs for each
of the experiments. To allow a fair comparison, the removed items in the test
set remain the same for all models during one run with a fixed pruning
parameter.

\paragraph{Results}
\begin{figure*}
  \includegraphics[width=\textwidth]{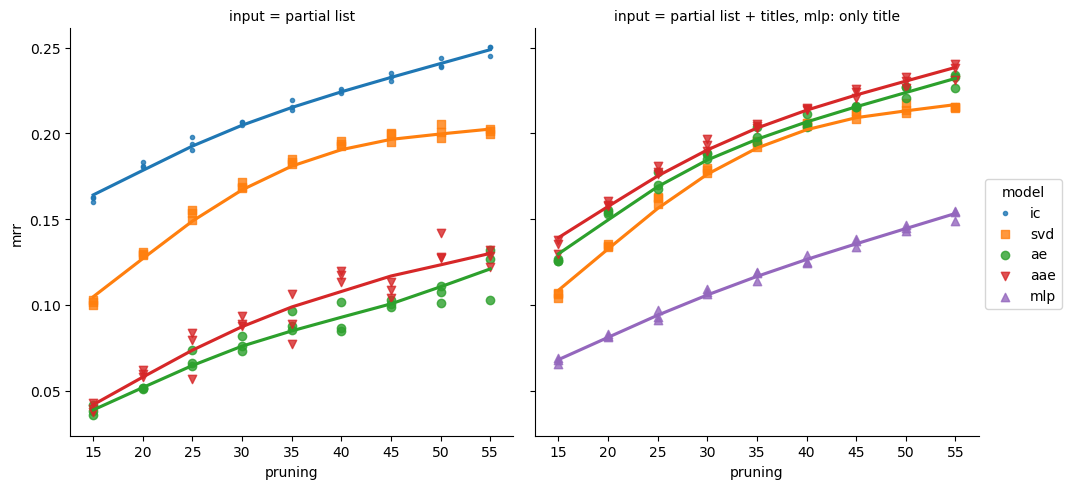}
  \caption{Mean reciprocal rank of missing citation on the test set with varying
  minimum item occurence (pruning) thresholds. Left: Only the partial list of items is given.
  Right: The partial list of items along with the document title is
given, except for the MLP, which can only make use of the title.}\label{fig:pubmed:results}
\end{figure*}

Figure~\ref{fig:pubmed:results} shows the results for the models with respect to
the pruning parameter that controls the number of considered items as well as
the sparsity (see Table~\ref{tab:pubmed:pruning}). We observe a trend that a more
aggressive pruning threshold leads to higher scores among all models. When no
title information is given, the item co-occurrence approach consistently
yields the highest scores. When title information is available,
adversarial autoencoders become competitive to the item co-occurrence approach
and yield higher scores than all of their components.

\subsection{Subject Label Recommendation}\label{sub:subjects}

On the basis of our experience in multi-label
classification~\cite{DBLP:conf/kcap/Grosse-BoltingN15,DBLP:conf/kcap/GalkeMSBS17},
we now consider a subject label recommendation task, which is close to how
professional subject indexers work.

\begin{figure}[!ht]
  \includegraphics[width=0.8\columnwidth]{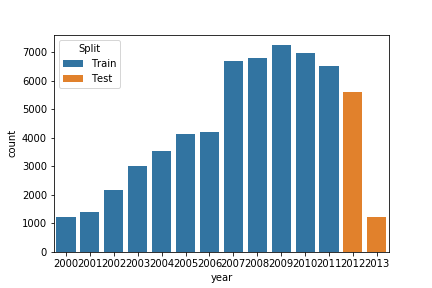}
  \caption{Count of documents by publication year starting with 2000 along with
  the split in training and test set for the Economics subject label dataset.}\label{fig:econ_by_year}
\end{figure}

\paragraph{Dataset}
The EconBiz dataset provided by ZBW --- Leibniz Information Centre for Economics
consists of 61,619 documents with label annotations from professional subject
indexers~\cite{DBLP:conf/kcap/Grosse-BoltingN15,DBLP:conf/kcap/GalkeMSBS17}. The
4,669 assigned labels are a subset of the controlled vocabulary Standardthesaurus
Wirtschaft\footnote{\url{http://zbw.eu/stw/version/latest/about}}.
The number of documents to which a label is assigned ranges between 1 and 13,925
with mean 69 (SD\@: 316) and median 14. The label annotations of a
document ranges between 1 and 23 with mean 5.24 (SD\@: 1.83) and median 5
labels.

\begin{table}
  \centering
  \caption{Dataset characteristics with respect to pruning thresholds on minimum item
  occurrence for the EconBiz subject label recommendation
  task.}\label{tab:economics:pruning}
  \begin{tabular}{lrrrr}
\toprule
pruning &  labels &  assigned labels &  documents &   density \\
\midrule
1  &     4,568 &     323,670 &    61,104 &  0.001160 \\
2  &     4,103 &     323,060 &    61,090 &  0.001289 \\
3  &     3,760 &     322,199 &    61,060 &  0.001403 \\
4  &     3,497 &     321,213 &    61,039 &  0.001505 \\
5  &     3,259 &     320,048 &    60,983 &  0.001610 \\
10 &     2,597 &     314,738 &    60,778 &  0.001994 \\
15 &     2,192 &     309,101 &    60,524 &  0.002330 \\
20 &     1,924 &     303,693 &    60,272 &  0.002619 \\
\bottomrule
\end{tabular}

\end{table}

\paragraph{Evaluation}
The preprocessing steps and evaluation procedure for the subject label
recommendation task is the same as in Section~\ref{sub:citations}. We also
conduct the split between training set and test set on the time axis (see
Figure~\ref{fig:econ_by_year}). This is challenging because label annotations suffer
from concept drift over time~\cite{DBLP:conf/jcdl/ToepferS17}. We use the years 2012 and 2013 as test
documents to obtain a train-test ratio similar to the scenario in
Section~\ref{sub:citations}. The dataset characteristics affected by dataset pruning are given in
Table~\ref{tab:economics:pruning}.

\paragraph{Results}

\begin{figure*}
  \includegraphics[width=\textwidth]{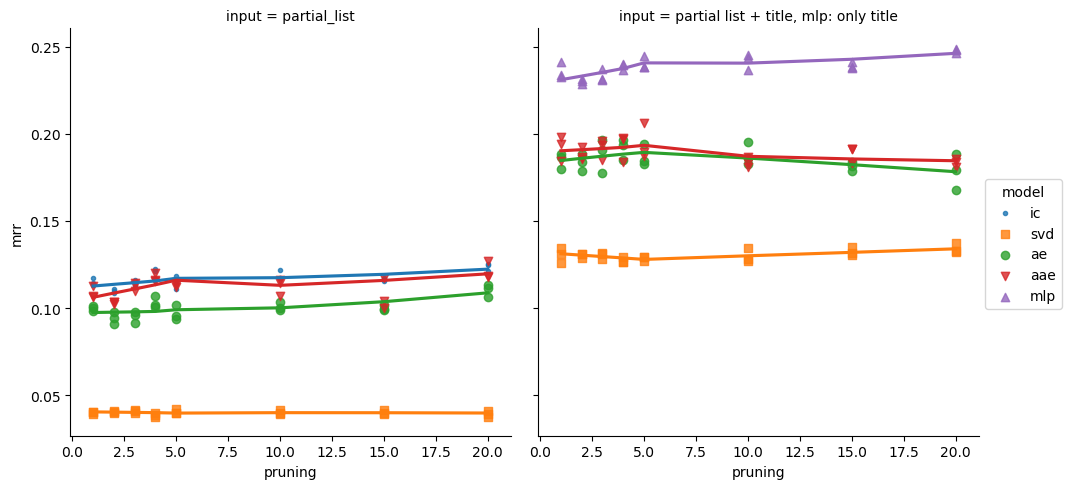}
  \caption{Mean reciprocal rank of missing subject label on the test set with varying
  minimum item occurence thresholds. Left: Only the partial list of items is given.
  Right: The partial list of items along with the document title is given, except
  for the MLP, which can only make use of the title.}\label{fig:economics:results}
\end{figure*}

Figure~\ref{fig:economics:results} shows the results for the models with respect
to the pruning parameter that controls the number of considered items and
therefore also the sparsity (see Table~\ref{tab:economics:pruning}). When no
title information is available, the adversarial autoencoder is
competitive to the item co-occurrence approach. When title information is
given, the adversarial autoencoder yields considerably higher scores than all
models operating without this information. The sole decoder part (an MLP-2
module) of the adversarial autoencoder yields, however, consistently higher
scores than the model as a whole.

\section{Discussion}\label{sec:discussion} 
We have evaluated adversarial autoencoders for two different recommendation tasks on scientific
documents with varying input modalities and varying numbers of considered items.
Our results reveal relationships between the type of recommendation task and the
input modalities. On the citation task, the partial list of citations is
relevant to recommend potentially missing citations. For the subject label
recommendation task, however, using solely the decoder on the title
information yields even better performance than the whole model.
Thus, our experiments show in which cases adversarial
autoencoders are beneficial. On the citation recommendation task
the title information enables adversarial autoencoders to become competitive to
the strong baseline from co-citation analysis. The effect of the adversarial
regularization component is marginal, yet leads to a consistent improvement over
traditional, undercomplete autoencoders. By imposing different thresholds on
minimum item occurrence, we varied the number of considered items and thus, the
degree of sparsity. We observe that all considered models are similarly affected
by the increased difficulty caused by higher numbers of considered items,
despite the high amount of parameters.

Even though it is not surprising that co-citation count is highly relevant for
citation recommendation~\cite{DBLP:journals/jasis/Small73}, we have shown that
adversarial autoencoders have a conceptual benefit: they offer the capability of
exploiting additional information along with the partial list of citations.
From the perspective of the model, it is of high importance to learn about the
prior distribution of the data, which explains the strength of the item
co-occurrence baseline. Autoencoders retain this benefit and may learn to put
appropriate weights in the bias parameters if it is helpful for the overall objective.
We envision that further types of information, such as the authors and
publication year may further increase the overall performance.

Compared to item co-occurrence or singular value decomposition, all
neural network approaches have a large number of learnable parameters as well as
hyperparameters that require tuning. To assess the quality of the model itself,
we used a fixed set of hyperparameters across all experiments and conducted
multiple runs of the same experimental setup to alleviate random effects in
initialization and shuffling.

On the subject recommendation task, we observed
that the MLP decoder alone yields higher mean reciprocal rank scores than the adversarial
autoencoder.  Thus, already assigned subjects are less informative for a subject
recommendation task than the titles are. This can be explained by a specific
guideline for subject indexers working on the specific EconBiz dataset that we
used for our experiments: when two or more subjects with a common ancestor in
the hierarchical thesaurus of subjects match, it is preferred to assign the
ancestor instead of the child subjects~\cite{DBLP:conf/kcap/Grosse-BoltingN15}.
Thus, two subjects that are semantically related because they share a common
ancestor are, because of the guideline, unlikely to co-occur in the annotations
of a single document.

We conducted \Nexperiments{} experiments over two different recommendation
tasks with different input modalities and varying degrees of sparsity.
While it is a limitation that we only use one dataset per task,
this enabled us to investigate the interactions across tasks, input modalities
and the effect of sparsity.
As a result, we can state that, on the one hand, there are tasks in which
co-occurrence implies relatedness. On the other hand, there are recommendation
tasks, in which co-occurrence of items rather implies diversity.

In the present work, we used one prototypical task for
each of these two types of recommendations, \ie{}, citations, where it is known that
co-citation reflects relatedness of the cited
resources~\cite{DBLP:journals/jasis/Small73,beel2016paper}, and subject labels,
where the guidelines of subject indexers suggest that
semantically related subjects are less likely to co-occur.
We have carefully
investigated the interaction between the semantics of item co-occurrence and
supplying the partial list of items as input for a recommender system.

For practical recommender systems, the present work offers evidence that the
aforementioned semantics of item co-occurrence is relevant for the decision,
whether the partial list of items should be supplied to a recommendation model
as input. We have shown that also on recommendation tasks, adversarial
autoencoders consistently outperform their traditional, undercomplete
counterpart and how additional information can be incorporated in such models.
Our results show that both models with no learnable parameters and models with a
high amount of learnable parameters are equally sensitive to the number of
considered items, which we controlled by pruning the datasets with respect to
minimum item occurrence.

\section{Conclusion}\label{sec:conclusion}

We conclude that the different semantic interpretation of item co-occurrence in
recommendation tasks highly affects the preferable input modalities. When item
co-occurrence resembles relatedness such as in citations, supplying the list of
already cited documents is beneficial for the overall performance. For subject
recommendations, we observe that co-occurring subjects does not imply that these
subjects are semantically similar. Rather, the document's subject needs to be
described by multiple, diverse subject annotations. In such cases, we have shown
that a single multi-layer perceptron component that operates only on the
documents' titles is stronger than the whole adversarial autoencoder. We have
shown that adversarial autoencoders consistently outperform undercomplete
autoencoders, and that their capability of incorporating multiple input
modalities offers a conceptual benefit.

\paragraph{Reproducibility}
The source code for reproducing our experiments is openly available on
GitHub\footnote{\url{https://github.com/lgalke/aae-recommender}}.

\begin{acks}
  This work was supported by the \grantsponsor{}{German Research Foundation}{} under project number \grantnum{}{311018540} (Linked Open Citation Database)
  as well as by the \grantsponsor{}{EU H2020}{} project MOVING (contract no \grantnum{}{693092}).
\end{acks}

\bibliographystyle{ACM-Reference-Format}
\balance{}
\bibliography{sigproc}

\end{document}